\documentclass[conference]{IEEEtran}
\IEEEoverridecommandlockouts
\usepackage{cite}
\usepackage{amsmath,amssymb,amsfonts}
\usepackage{algorithmicx}
\usepackage{graphicx}
\usepackage{textcomp}
\usepackage{xcolor}
\usepackage{flushend}
\usepackage{float}
\usepackage{url}
\usepackage{algorithm}
\usepackage{algpseudocode}
\usepackage{tikz}
\usepackage{pgfplots}
\pgfplotsset{compat=1.18}
\usepackage{multirow}
\usepackage{graphicx}
\usepackage{subcaption}
\def\BibTeX{{\rm B\kern-.05em{\sc i\kern-.025em b}\kern-.08em
    T\kern-.1667em\lower.7ex\hbox{E}\kern-.125emX}}
\begin{document}

\newcommand{\etal}{\textit{et al}.}

\title{Full-Frequency Temporal Patching and Structured Masking for Enhanced Audio Classification}

\author{Aditya Makineni, Baocheng Geng, Qing Tian\\
Department of Computer Science, University of Alabama at Birmingham\\
Birmingham, Alabama, USA \\
\{amakinen,bgeng,qtian\}@uab.edu
\thanks{*This work was partially supported by the National Science Foundation (NSF) under Award No. 2412285. (Corresponding author: Qing Tian.)}% <-this % stops a space
%\and
%\IEEEauthorblockN{Baocheng Geng}
%\IEEEauthorblockA{\textit{Department of Computer Science} \\
%\textit{University of Alabama at Birmingham}\\
%Birmingham, Alabama \\
%bgeng@uab.edu}
%\and
%\IEEEauthorblockN{Qing Tian}
%\IEEEauthorblockA{\textit{Department of Computer Science} \\
%\textit{University of Alabama at Birmingham}\\
%Birmingham, Alabama \\
%qtian@uab.edu}
}

\maketitle

\begin{abstract}
Transformers and State-Space Models (SSMs) have advanced audio classification by modeling spectrograms as sequences of patches. However, existing models such as the Audio Spectrogram Transformer (AST) and Audio Mamba (AuM) adopt square patching from computer vision, which disrupts continuous frequency patterns and produces an excessive number of patches, slowing training, and increasing computation. We propose Full-Frequency Temporal Patching (\textit{FFTP}), a patching strategy that better matches the time-frequency asymmetry of spectrograms by spanning full frequency bands with localized temporal context, preserving harmonic structure, and significantly reducing patch count and computation. We also introduce \textit{SpecMask}, a patch-aligned spectrogram augmentation that combines full-frequency and localized time-frequency masks under a fixed masking budget, enhancing temporal robustness while preserving spectral continuity. When applied on both AST and AuM, our patching method with SpecMask improves mAP by up to \textit{+6.76} on AudioSet-18k and accuracy by up to \textit{+8.46} on SpeechCommandsV2, while reducing computation by up to \textit{83.26\%}, demonstrating both performance and efficiency gains.
\end{abstract}

\section{Introduction}
Recent advances in deep learning for audio classification have been driven by architectures originally developed for other modalities, particularly Transformers \cite{vaswani2017attention} and State Space Models (SSMs) \cite{gu2021efficiently}. These models capture long-range dependencies of audio spectrogram patches, achieving state-of-the-art performance on large-scale benchmarks. 

% Audio Spectrogram Transformer (AST) \cite{gong2021ast} introduced the direct adaptation of Vision Transformers (ViTs) \cite{dosovitskiy2020image} to audio classification, treating log-mel spectrograms as 2D images and partitioning them into fixed-size square patches. These patches are linearly embedded before being fed into a standard Transformer encoder. However, the square patching strategy, inherited from ViTs, overlooks the asymmetric nature of spectrograms. It imposes the same resolution along both axes, which may not align well with the distinct temporal and spectral characteristics of audio signals.
% %In audio, frequency resolution is fixed by the Short-Time Fourier Transform (STFT), and harmonic structures extend continuously across the full frequency axis. As a result, 
% Arbitrary square patching may disrupt critical frequency patterns and continuity, while producing an excessive number of patches. This excessive patch proliferation increases memory usage, training time, and computational cost without providing proportional gains in performance.
% Most recently, Audio Mamba (AuM) \cite{erol2024audio} adapted the Mamba SSM architecture \cite{gu2023mamba} to spectrogram sequences, leveraging the long-context modeling and linear-time scaling of SSMs. Although AuM reduces the computational complexity of modeling long sequences, it retains the same square patching design as AST, and thus inherits its inefficiencies and structural misalignment with spectrograms

Audio Spectrogram Transformer (AST) \cite{gong2021ast} and the more recent Audio Mamba (AuM) \cite{erol2024audio} both adapt image-based architectures to audio classification by treating log-mel spectrograms as 2D images and partitioning them into fixed-size square patches, which are then linearly embedded and processed by their respective sequence models. In AST, these patches are passed to a standard Transformer encoder, while AuM replaces the Transformer with the Mamba SSM architecture to enable long-context modeling with linear-time scaling. However, the square patching strategy, borrowed from Vision Transformers (ViTs) \cite{dosovitskiy2020image}, overlooks the asymmetric nature of spectrograms, imposing the same resolution along both axes despite their distinct temporal and spectral characteristics. This design can disrupt critical frequency patterns and continuity while producing an excessive number of patches, increasing memory usage, training time, and computational cost without proportional performance gains. Since AuM retains the same square patching design as AST, it inherits such inefficiencies and structural misalignments with spectrograms.

To address these limitations, we propose Full-Frequency Temporal Patching (FFTP), a patching strategy tailored to the time-frequency characteristics of spectrograms. In our method, patches span the full frequency range while capturing localized temporal context. This preserves harmonic continuity and significantly reduces the number of patches compared to square patching. As a result, it improves the efficiency of both Transformer and SSM based architectures while aligning the model's receptive field with the natural structure of spectrograms.

To further improve temporal robustness, we introduce SpecMask, a patch-aligned spectrogram masking method tailored to FFTP. SpecMask combines full-frequency time masks with smaller, localized time-frequency masks under a fixed masking budget. This improves temporal robustness while preserving spectral coherence. By aligning augmentation masks to patch boundaries, SpecMask operates at the same granularity as the model's input tokens, enhancing regularization effectiveness while preserving spectral coherence.

Our experiments on two widely used audio classification benchmarks, AudioSet-18k~\cite{gemmeke2017audio} and SpeechCommandsV2~\cite{warden2018speech}, demonstrate that combining FFTP with SpecMask yields consistent improvements in both accuracy and mean average precision (mAP) while significantly reducing computational cost. These results highlight the importance of designing input representations and augmentations that are structurally aligned with the properties of audio spectrograms.

% We summarize the contributions of this work: (1) A rectangular patching strategy for audio spectrograms that preserves full frequency information and reduces token count quadratically. (2) SpecMask, a spectrogram augmentation method aligned with rectangular patches to improve temporal robustness. (3) Consistent performance gains on both Transformer and SSM architectures, demonstrating the generality of the proposed design.

% The rectangular design also brings significant efficiency gains: compared to square patching, our method reduces the number of patches from 1212 to 196, cuts training time from 6.5 to 2.5 hours, lowers GFLOPs from 103.25 to 17.30, and reduces inference latency from 14.50 to 2.18 ms per sample, while still achieving higher performance.

\section{Related Work}
%\subsection{Different Kernels shapes in CNNs}
While square patches are commonly used in spectrogram processing, only a few prior studies have explored the use of various kernel shapes and patches for audio spectrogram analysis, and most of these focus on CNNs. Research by Pons \etal \cite{pons2019randomly} demonstrated that the use of vertical and horizontal filters in CNNs for music audio classification could capture frequency and temporal patterns more effectively than square kernels. Their work showed that combining these specialized filters improved performance in different music information retrieval tasks. In the field of speech recognition, Abdel-Hamid \etal \cite{abdel2014convolutional} proposed using limited weight sharing in CNN architectures, effectively creating rectangular receptive fields that were better suited for capturing local spectro-temporal patterns in speech spectrograms. This approach led to improved performance on phoneme recognition tasks compared to conventional CNNs with square filters. In the context of environmental sound classification, Piczak \cite{piczak2015environmental} experimented with various CNN architectures and found that rectangular filters performed better than square filters, especially when aligned with the time axis of the spectrogram. While these studies highlight the benefits of anisotropic receptive fields, they operate within convolutional frameworks, where kernels slide locally across the input. In contrast, our approach applies full frequency patches in Transformer and SSM architectures, where each patch serves as a global token in a sequence model, enabling long-range modeling across the entire frequency axis rather  than local convolutional aggregastion.
% These results motivate our use of full-frequency rectangular patches in Transformer and SSM architectures.

%\subsection{Masking-based approaches}
Spectrogram masking is essential for regularizing audio models. SpecAugment \cite{park2019specaugment} applies random time and frequency masks, with variants for dynamic sizing~\cite{lu2024sample} and mask scheduling~\cite{byun2023effective}. However, these approaches ignore the patching structure of downstream models. The proposed SpecMask aligns masks with FFTP, combining full-band temporal masks with smaller localized time-frequency masks. This preserves spectral coherence while improving temporal robustness, making it well-suited for patch-aligned architectures.

\section{Methodology}\label{sec:methodology}
\begin{figure}[h]
    \centering
    \includegraphics[width=1.0\linewidth]{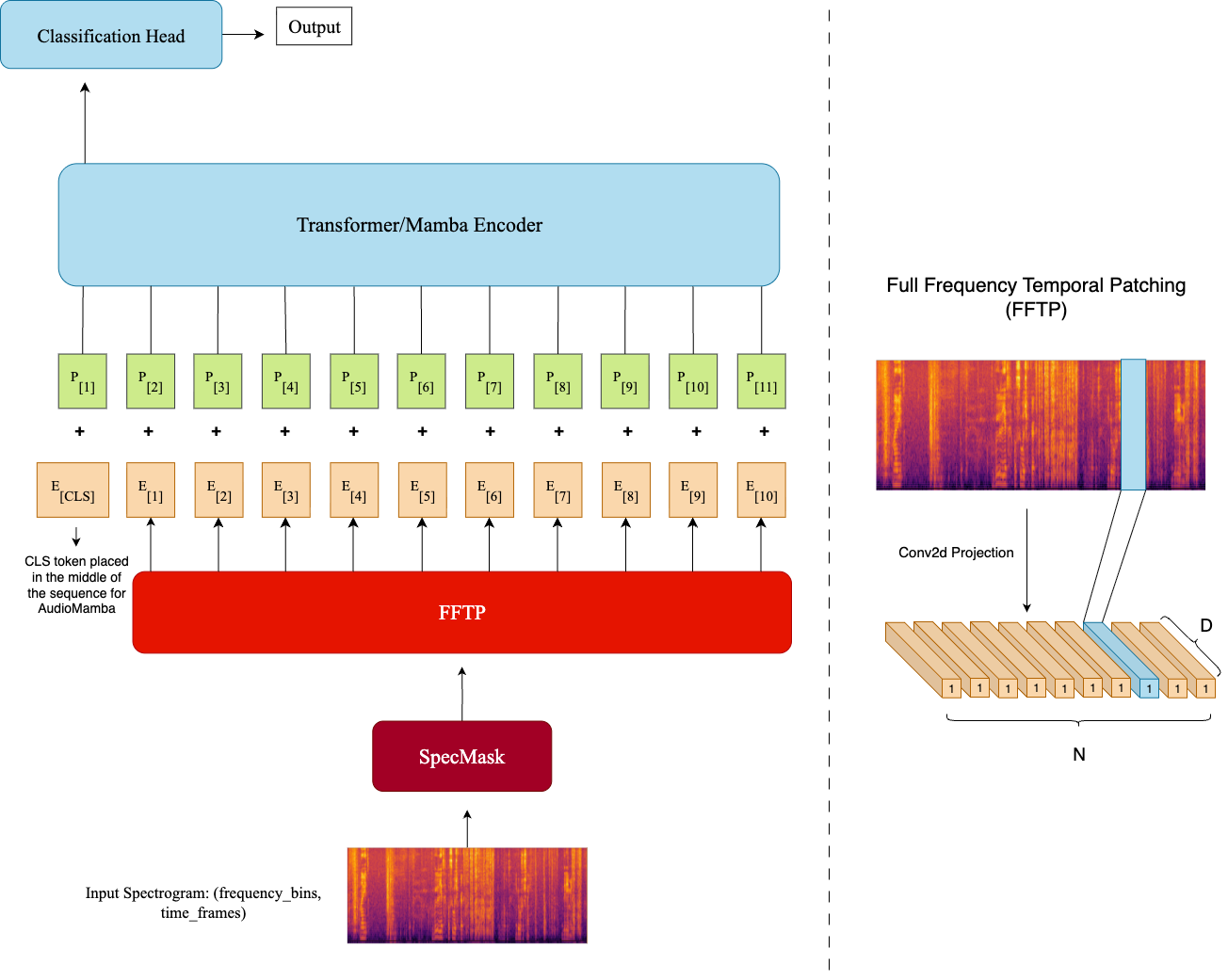}
    \caption{Architectures of models trained and an illustration of Full-Frequency Temporal Patching: A log-mel spectrogram is projected into a sequence of $D$-dimensional embeddings using a 2D Convolution layer with kernal size $F, T_p$ and stride $F, s_t$. Each patch spans the full frequency axis while capturing a short temporal window.}
    \label{fig:Model Architecture}
\end{figure}

To show the effectiveness of the proposed FFTP and SpecMask methods, we conduct experiments using two sequence-based architectures for audio classification: Audio Spectrogram Transformer (AST) and Audio Mamba (AuM). Figure \ref{fig:Model Architecture} illustrates the model architectures along with integration of FFTP into the patch embedding stage.

\subsection{Full-Frequency Temporal Patching (FFTP)}
In this paper, we propose Full-Frequency Temporal Patching (FFTP), a patching strategy that better aligns with the time-frequency asymmetry of audio spectrogram. Unlike conventional square patching, our method decouples the patch dimensions along the time and frequency axes, allowing each patch to span the full frequency range while capturing localized temporal context. Beyond improving efficiency, this patching method preserves harmonic and spectral structures, such as formants and harmonics, that extend across frequency bins, resulting in semantically richer and more coherent token representations.

Specifically, the input waveform is first converted to a mono signal and uniformly sampled. A log-mel spectrogram $X \in \mathbb{R}^{B \times 1 \times F \times T}$ is then computed, where $B$ is the batch size, $F$ is the number of mel-frequency bins (e.g., 128), and $T$ is the number of time frames.

To extract embeddings, we apply a 2D convolutional layer with kernel size $(F_p, T_p)$ and stride $s = (s_f, s_t)$:

\[
Z = \text{Conv2D}(X; W_c, s) \in \mathbb{R}^{B \times D \times 1 \times N}\,,
\]

\noindent where $W_c \in \mathbb{R}^{D \times 1 \times F_p \times T_p}$ is the learnable convolutional kernel, $(F_p, T_p)$ is the patch size, $D$ denotes the embedding dimension, and $N = \left\lfloor \frac{T - T_p}{s_t} + 1 \right\rfloor$ is the number of temporal patches.

In our configuration, $F_p = F$ and $s_f = F$, meaning each patch spans the entire frequency axis with no overlap in frequency. The temporal stride $s_t$ controls the degree of overlap in time, allowing flexible temporal resolution.

% \begin{figure}[h]
%     \centering
%     \includegraphics[width=0.95\linewidth]{Conv2d_Proj_diagram.png}
%     \caption{Illustration of Full-Frequency Temporal Patching: A log-mel spectrogram is projected into a sequence of $D$-dimensional embeddings using a 2D Convolution layer with kernal size $F, T_p$ and stride $F, s_t$. Each patch spanns the full frequency axis while capturing a short temporal window.}
%     \label{fig:Conv2d}
% \end{figure}

The output $Z$ contains $D$-dimensional embeddings for each of the $N$ time-localized patches, where the original frequency dimension $F$ has been projected into the embedding space. The result is then reshaped into a sequence of token embeddings:
% with the frequency dimension collapsed to 1
\[
Z' = \text{Transpose}(\text{Flatten}(Z)) \in \mathbb{R}^{B \times N \times D}\,,
\]
\noindent where each D-dimensional row represents the patch embedding of an audio sample at a specific time.

This procedure is illustrated in Figure \ref{fig:Model Architecture} where the spectrogram is transformed into a sequence of tall, narrow patches, each encoding a short time window with complete frequency coverage. This stands in contrast to square patching, which slices the spectrogram into small, spectrally constrained fragments, disrupting the continuity of important frequency patterns.

\begin{algorithm}[h]
\caption{Proposed SpecMask Algorithm}
\label{alg:SpecMask_Alg}
\begin{algorithmic}[1]
\State \textbf{Input:} Spectrogram $X \in \mathbb{R}^{H \times W}$, masking budget $A$, maximum patch size $(\text{max\_h}, \text{max\_w})$, mask type $(\text{mask\_value})$
\State \textbf{Output:} Masked spectrogram $X'$
\State $M \gets 0_{H \times W}$ \Comment{Mask map}
\State $masked\_area \gets 0$
\If{$\text{mask\_value} = \text{mean}$} 
    \State $\mu \gets \text{mean}(X)$
\EndIf
\While{$masked\_area < A$}
    \If{random() $< 0.7$} 
        \State $h \gets H$ \Comment{Full-frequency patch}
        \State $w \gets$ random width $\le$ max\_w
    \Else
        \State $h \gets$ random height $\le$ max\_h
        \State $w \gets$ random width $\le$ max\_w
    \EndIf
    \State choose random $(x,y)$ where $M[x:x+h, y:y+w] = 0$
    \State apply mask to $X[x:x+h, y:y+w]$ using $\text{mask\_value}$
    \State $M[x:x+h, y:y+w] \gets 1$
    \State $masked\_area \gets masked\_area + h*w$
\EndWhile
\State \textbf{return} $X'$
\end{algorithmic}
\end{algorithm}

\subsection{SpecMask: Patch-Aligned Spectrogram Masking}
\label{specMask}

To improve the generalization of models under FFTP, we introduce \textit{SpecMask} (Algorithm \ref{alg:SpecMask_Alg}), a spectrogram masking strategy designed to align the structure of masking with the geometry of the input spectrogram.

\begin{figure}[t]
    \centering
    \begin{subfigure}[b]{0.45\linewidth}
        \centering
        \includegraphics[width=\columnwidth]{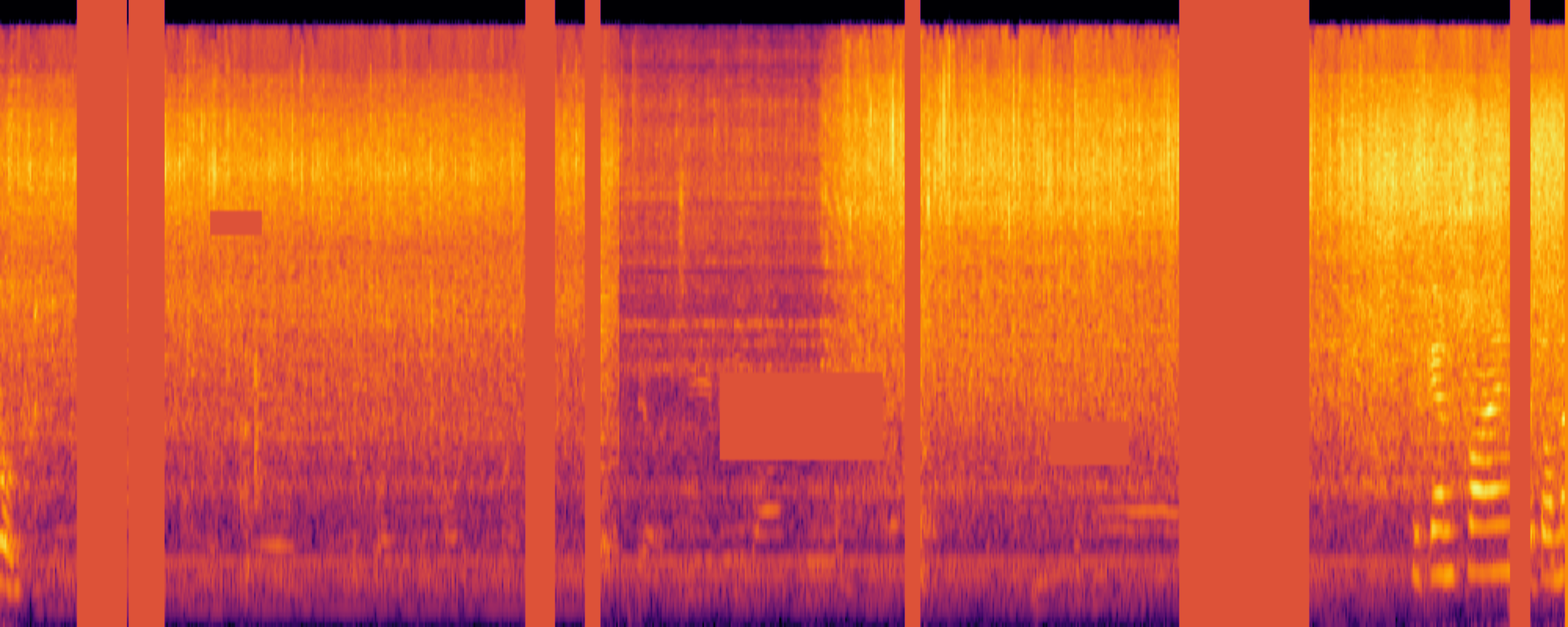}
        \caption{SpecMask}
        \label{fig:specMask_illustration}
    \end{subfigure}
    \begin{subfigure}[b]{0.45\linewidth}
        \centering
        \includegraphics[width=\columnwidth]{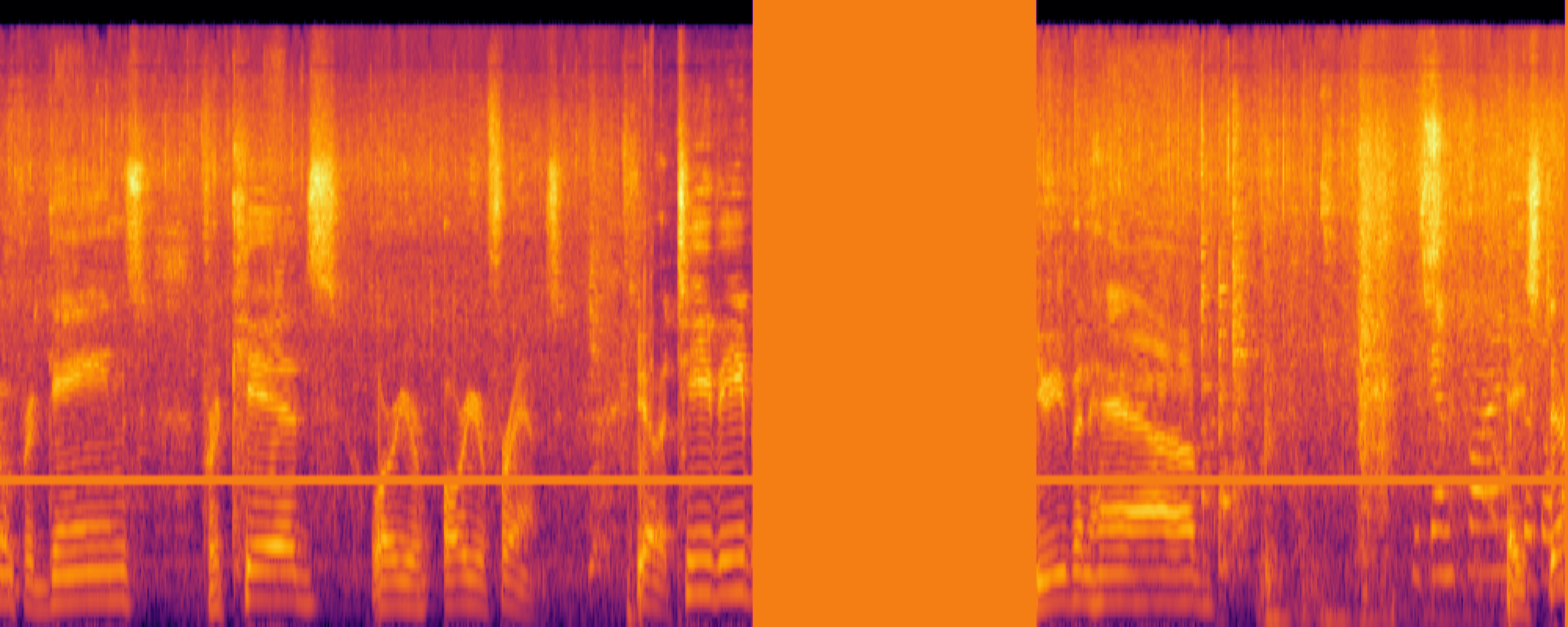}
        \caption{SpecAugment}
        \label{fig:spegAug_illustration}
    \end{subfigure}
    \caption{Visual differences between the proposed SpecMask and standard SpecAugment}
    \label{fig:Augmentation_comparision}
\end{figure}

While full frequency temporal masking is present in standard SpecAugment, SpecMask enforces a structured, patch-aligned masking strategy that prioritizes semantically coherent corruptions. In our case, 70\% of masked area consists of full temporal masks that aligns with the model's receptive fields, while the remaining 30\% uses smaller, localized time-frequency masks to maintain diversity as seen in Figure \ref{fig:Augmentation_comparision}. This controlled balance ensures that the model learns to rely on global spectral structure while being exposed to realistic temporal gaps.

Masks are applied without overlap under a fixed area budget (e.g., 20\% of the spectrogram), with up to 100 placement attempts per mask to avoid clustering. Masked regions are filled with the spectrogram mean to reduce bias.

By matching the augmentation strategy to the patch layout, SpecMask enhances regularization and improves temporal robustness in a way that is consistent with the model's input structure.

\section{Experiments and Results}
\subsection{Experimental Setup}
%We conducted experiments to evaluate the impact of rectangular patching on audio classification performance and efficiency. 
To demonstrate the effectiveness of our FFTP and SpecMask, we trained two state-of-the-art audio models as shown in Figure \ref{fig:Model Architecture}, AST and AuM, from scratch on two benchmark datasets: Audioset(balanced subset)~\cite{gemmeke2017audio} and SpeechCommandsV2~\cite{warden2018speech}. All experiments were run on a single NVIDIA A100 GPU with 80 GB of VRAM.

The AudioSet-18k contains 527 multi-label audio classes, with each sample approximately 10 seconds long. Due to the unavailable or restricted content (e.g., deleted videos, private accounts), we successfully retrieved 18,684 out of the original 22,176 samples. We used a pre-downloaded version made publicly available via Hugging Face \cite{agkphysicsAudioSetDataset}. SpeechCommandsV2 is a single-label dataset consisting of around 65,000 one-second utterances across 35 spoken words.

All audio was converted to mono channel and uniformly sampled at 16 kHz. For AudioSet, we first perform mixup~\cite{zhang2017mixup} on the raw waveforms with interpolation ratio of 0.5 before transforming them into log-mel spectrograms of size \textit{128$\times$1000} (frequency bins x time-frames). For SpeechCommandsV2, spectrograms were resized to \textit{128$\times$128}. Mel-filter banks were computed using standard setting consistent with the original implementations of AST and AuM, with torchaudio kaldi fbank parameters: $htk\_compact = True$, $window\_type = 'hanning'$, and $frame\_shift=10$.

%We applied consistent pre-processing and augmentation pipelines across both models for each dataset. For the baseline models with square patching, we used standard SpecAugment with settings consistent with the best configuration of AST and AuM.
%For rectangular patching we adjusted the augmentation strategy to better match the full frequency structure. 
We conducted experiments using both SpecAugment and our proposed SpecMask. When using SpecAugment with FFTP, we applied time masking with a maximum of 400 time frames and frequency masking with a maximum of 5 bins for AudioSet and a maximum of 15 time frames and 5 bins for SpeechCommandsV2. This is done to prevent over corruption of frequency bands in FFTP. When using the proposed SpecMask, we set the total mask area to 25,600 with maximum height 128 and maximum width 128 for AudioSet and to 1,024 with maximum height 128 and maximum width 16 for SpeechCommandsV2. In all SpecMask cases, the masked regions were filled with the spectrogram mean.

% Training was performed as follows:
% \begin{itemize}
%     \item \textbf{AudioSet}: \\
%     Batch size = 12, 25 epochs. \\
%     Optimizer: AdamW with learning rate $5 \times 10^{-5}$, weight decay $5 \times 10^{-7}$, betas $(0.95, 0.999)$. \\
%     Learning rate schedule: 5-epoch linear warm-up followed by cosine decay. \\
%     Loss: Binary cross-entropy with logits (BCEWithLogitsLoss).
%     \item \textbf{SpeechCommandsV2}: \\
%     Batch size = 256, 20 epochs. \\
%     Optimizer: AdamW with learning rate $1 \times 10^{-3}$, weight decay $1 \times 10^{-6}$, betas $(0.95, 0.999)$. \\
%     Learning rate schedule: 3-epoch linear warm-up followed by cosine decay. \\
%     Loss: Cross-entropy.
% \end{itemize}

Models were trained on AudioSet-18k for 25 epochs with a batch size of 32 and on SpeechCommandsV2 for 20 epochs with a batch size of 256. We used the AdamW optimizer with a linear warm-up followed by cosine decay of the learning rate. The loss was binary cross-entropy for AudioSet and categorical cross-entropy for SpeechCommandsV2.

\subsection{Quantitative Results}
We evaluated performance using standard metrics such as mean average precision (mAP) for multi-label classification on AudioSet, and accuracy (Acc.) for single-label classification on SpeechCommandV2. All results are shown in Table \ref{tab:model_comparison}.

% Results table palced here for alignment
\begin{table*}[t]
    \centering
    \begin{tabular}{lccccc} 
        \hline
        Model & AudioSet-18K (mAP) & Speech Comm. V2 (Acc.)\\
        \hline
        AST Square & 11.25 & 85.27\\
        \textbf{AST with FFTP} & \textbf{15.38} & \textbf{93.73}\\
        \textbf{AST with FFTP + SpecMask} & \textbf{18.32} & \textbf{95.94}\\
        AuM Square & 13.28 & 91.58\\
        \textbf{AuM with FFTP} & \textbf{14.24} & \textbf{94.68}\\
        \textbf{AuM with FFTP + SpecMask} & \textbf{17.59} & \textbf{96.49}\\
        \hline
    \end{tabular}
    \vspace{0.5em}
    \caption{From-scratch training results. Models without “+ SpecMask” use SpecAugment, while “+ SpecMask” variants use our proposed SpecMask.}
    \label{tab:model_comparison}
\end{table*}

According to the results, our FFTP strategy consistently outperforms the conventional AST and AuM across all tested datasets, with its performance further enhanced by SpecMask. This advantage stems from the fact that, in a spectrogram, the time and frequency dimensions have distinct semantics and scales. By preserving the continuity of spectral patterns, FFTP introduces an inductive bias that aligns better with how signals vary over time and frequency (while also greatly reducing computation, as will be shown in Sec. \ref{sec:efficiency}).

\subsection{Attention Overlay Analysis}
\begin{figure}[t]
    \centering
    \begin{subfigure}[b]{0.9\linewidth}
        \centering
        \caption{Square Patch Attention}
        \includegraphics[width=0.9\columnwidth]
        {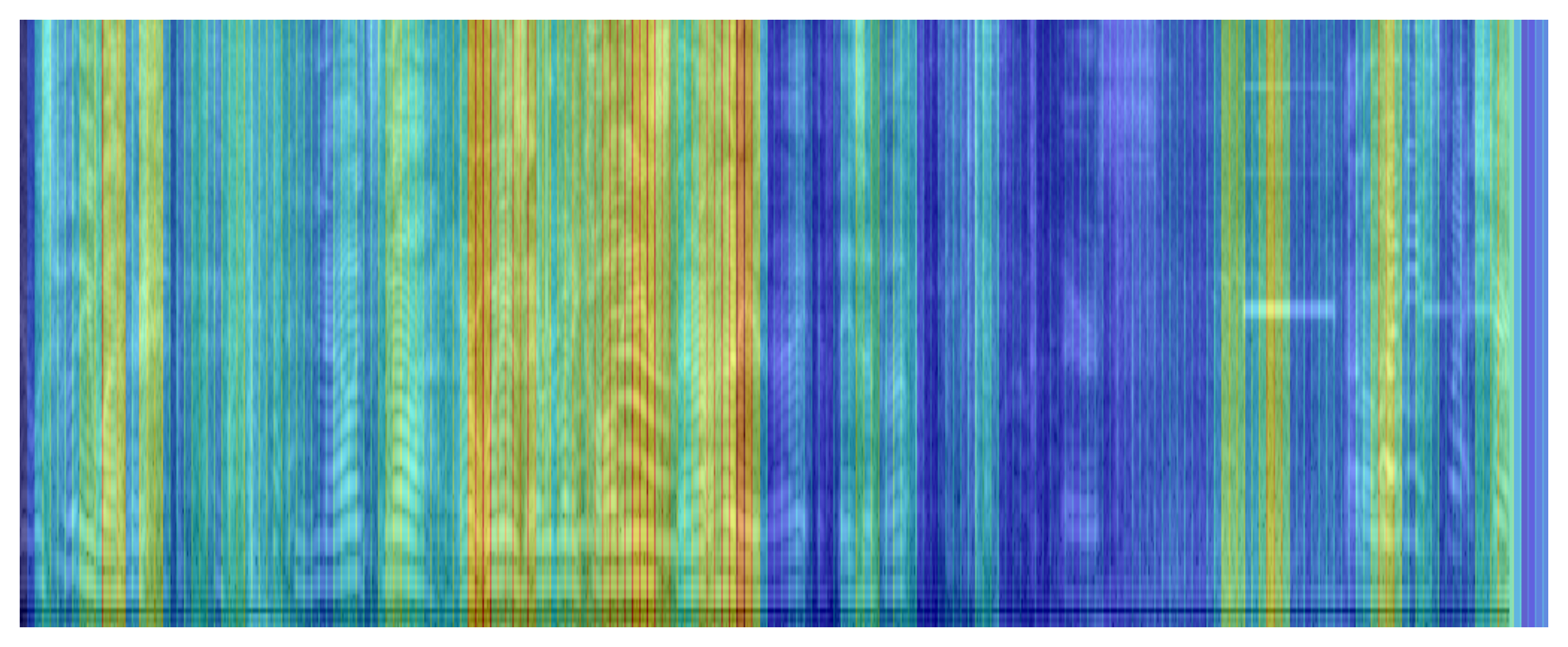}
        \label{fig:square_attn}
    \end{subfigure}

    \vspace{1em}
    
    \begin{subfigure}[b]{0.85\columnwidth}
        \centering
        \resizebox{\linewidth}{!}{%
        \begin{tikzpicture}
            % Timeline
            \draw[thick] (0,0) -- (10,0);
            
            % Tick marks at event times
            \draw (0,0.1) -- (0,-0.1);
            \draw (1,0.1) -- (1,-0.1);
            \draw (4,0.1) -- (4,-0.1);
            \draw (5,0.1) -- (5,-0.1);
            \draw (6,0.1) -- (6,-0.1);
            \draw (7,0.1) -- (7,-0.1);
            \draw (10,0.1) -- (10,-0.1);
    
            % Labels
            \node[below] at (0, -0.2) {0:00};
            \node[below] at (1, -0.2) {0:01};
            \node[below] at (4, -0.2) {0:04};
            \node[below] at (5, -0.2) {0:05};
            \node[below] at (6, -0.2) {0:06};
            \node[below] at (7, -0.2) {0:07};
            \node[below] at (10, -0.2) {0:10};
    
            % Event annotations above timeline
            \node[above] at (0.5,0.25) {Car Doors};
            \node[above] at (4.5,0.25) {Key Jangling};
            \node[above] at (6.5,0.25) {Car Engine};
        \end{tikzpicture}%
        }
        \label{fig:audio_timeline}
    \end{subfigure}
    
    \begin{subfigure}[b]{0.9\linewidth}
        \centering
        \includegraphics[width=0.9\columnwidth]{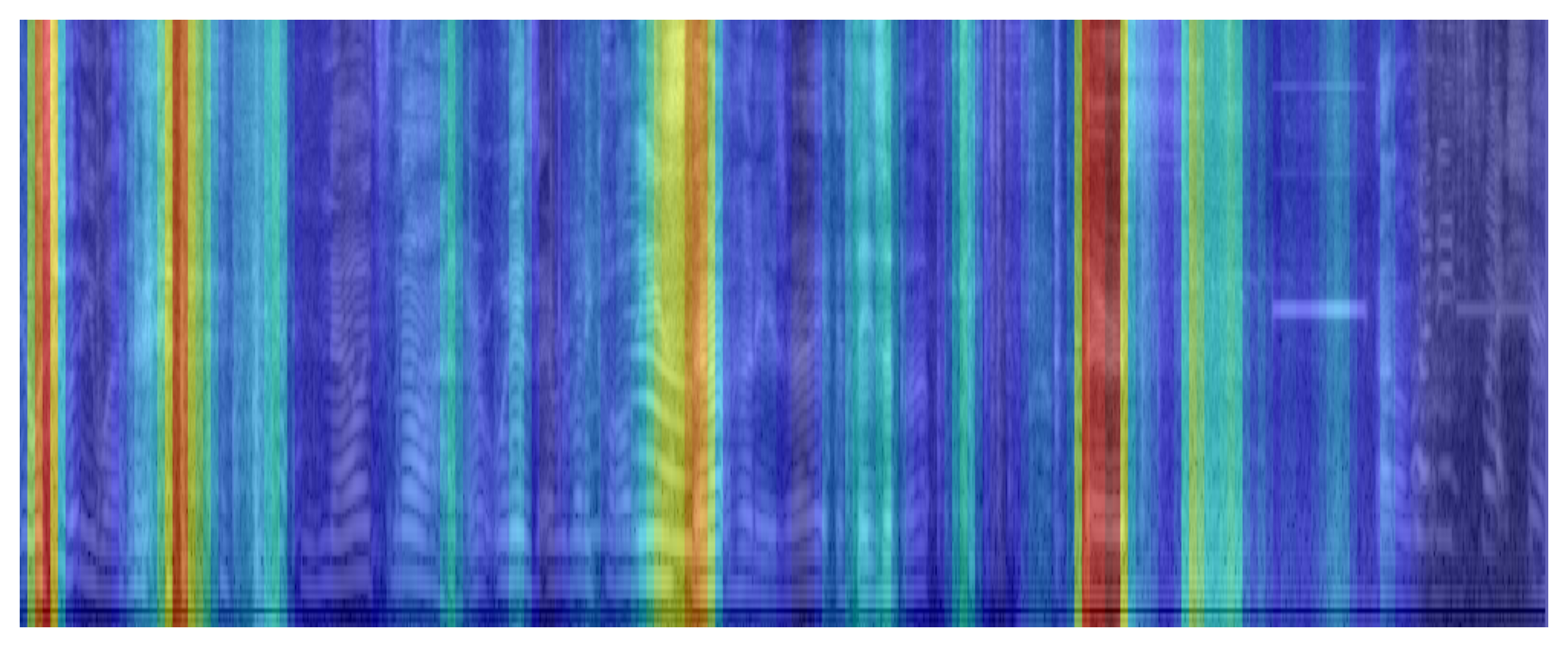}
        \caption{FFTP + SpecMask Attention}
        \label{fig:rect_attn}
    \end{subfigure}
    \caption{Attention maps of Baseline AST (Square Patch + SpecAug) and AST with FFTP + SpecMask on AudioSet-18k sample 0RWZT-miFs with labels ``Keys Jangling'' and ``Car''. The middle timeline shows annotated events; unmarked regions correspond to background noise.}
    %Attention maps of the Baseline AST (Square Patch) and AST with FFTP + SpecMask on a test sample from AudioSet-18k (ID: -0RWZT-miFs) with labels ``Keys Jangling'' and ``Car''. The timeline in the middle shows the annotated audio events within the 10-second clip; all other parts of the audio are considered background noise.
    \label{fig:Attn_comparision}
\end{figure}
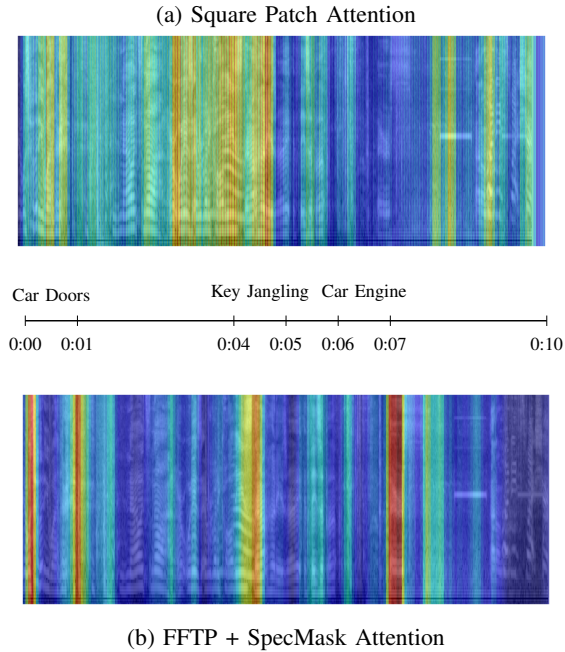

To demonstrate how the models focus on different parts of the spectrograms, we leverage Attention Rollout \cite{abnar2020quantifying}.
Figure \ref{fig:Attn_comparision} illustrates the attention of the AST baseline and our FFTP + SpecMask model overlaid on the same spectrogram. We can see that our FFTP + SpecMask model (Fig. \ref{fig:rect_attn}) tends to focus more on the high-energy regions of the spectrogram, such as distinct vertical and localized patterns, while effectively ignoring background noise. This precise localization suggests a more meaningful alignment with relevant acoustic events. In contrast, the square patch model (Fig. \ref{fig:square_attn}) exhibits broader attention coverage, capturing larger areas of the spectrogram, including regions that do not contain critical information, potentially diluting its sensitivity to critical features.
%This is because rectangular patches capture a more elongated view of the spectrogram that spans across the frequency axis, which allows the model to see a broader range of frequencies within the time frames in a given patch. This allows the model to focus on the full spectrum of features that correspond to the audio event, rather than being restricted to a smaller, more localized area.
This difference arises from the structure of the patches extracted using FFTP that span a wider frequency range within each time frame, allowing the model to capture more complete and coherent spectral patterns, such as harmonics, formants, or broadband events. By better aligning with the natural structure of spectrograms, FFTP provides a stronger inductive bias toward relevant frequency features, reduce fragmentation across patch boundaries, and support more context-aware attention. In contrast, square patches miss critical frequency components due to their limited coverage, breaking the inherent continuity of the spectrum.

%\subsection{Ablation Study}
%To better understand the influence of patch configuration and spectrogram-level augmentation, a series of ablation experiments were conducted using the AST model trained on the AudioSet balance subset.

\subsection{Patch Count and Efficiency Analysis}\label{sec:efficiency}

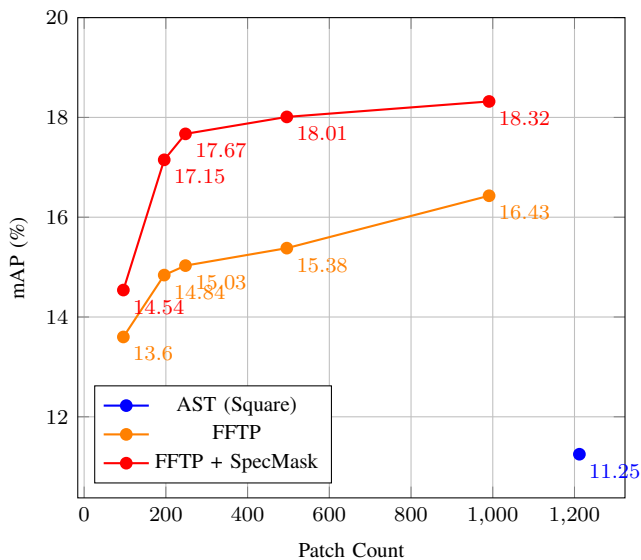
\begin{figure}[h]
    \centering
    \begin{tikzpicture}
        \begin{axis}[
            width=1.0\columnwidth,
            height=0.9\columnwidth,
            xlabel={Patch Count},
            ylabel={mAP (\%)},
            grid=major,
            ymax=20,
            legend pos=south west,
            tick label style={font=\footnotesize},
            label style={font=\footnotesize},
            legend style={font=\footnotesize},
            nodes near coords,
            every node near coord/.style={font=\footnotesize, below right}
        ]
        % AST (Square)
        \addplot[color=blue, mark=*, thick] coordinates {
            (1212, 11.25)
        };
        \addlegendentry{AST (Square)}

        % FFTP
        \addplot[color=orange, mark=*, thick]
        coordinates {
            (96, 13.60)
            (196, 14.84)
            (248, 15.03)
            (496, 15.38)
            (991, 16.43)
        };
        \addlegendentry{FFTP}

        % FFTP + SpecMask
        \addplot[color=red, mark=*, thick] coordinates {
            (96, 14.54)
            (196, 17.15)
            (248, 17.67)
            (496, 18.01)
            (991, 18.32)
        };
        \addlegendentry{FFTP + SpecMask}
        \end{axis}
    \end{tikzpicture}
    \caption{Patch count vs mAP for square patching and FFTP on AudioSet-18k.}
    \label{fig:patch_comparison}
\end{figure}

In attention-based transformers, the number of patches plays a crucial role in determining model efficiency. In this section, we first analyze the relationship between model performance and patch count by varying patch and stride settings using the AST model on the AudioSet-balanced dataset. 
Table~\ref{tab:patch_configs} reports how patch size and stride configurations translate into the number of extracted patches, while Figure~\ref{fig:patch_comparison} illustrates the relationship between patch count and classification performance (mAP).

\begin{table}[h]
    \centering
    \begin{tabular}{lccc} 
        \hline
        Patch Shape & Patch Size & Stride & Patches \\
        \hline
        Square & (16, 16) & (10, 10) & 1212 \\
        \hline
        \multirow{4}{*}{FFTP}
               & (128, 50) & (128, 10) & 96 \\
               & (128, 25) & (128, 5) & 196 \\
               & (128, 10) & (128, 4)  & 248 \\
               & (128, 10) & (128, 2)  & 496 \\
               & (128, 10) & (128, 1)  & 991 \\
        \hline
    \end{tabular}
    \vspace{0.5em}
    \caption{Patch configurations and resulting patch counts for AudioSet18k.}
    \label{tab:patch_configs}
\end{table}

As shown in Table \ref{tab:patch_configs}, reducing the stride increases the overlap between patches, resulting in a larger number of patches and a finer-grained temporal representation. Figure \ref{fig:patch_comparison} shows that higher patch counts produced with greater overlap steadily improved performance, with the best results achieved at the highest overlap setting (stride = 1) that produces 991 patches. 
% However, as illustrated in Figure \ref{fig:patch_comparison}, this does not translate into unlimited performance gains. While moderate overlap yields noticeable improvements in mAP, the benefit gradually plateaus, and at the extreme configuration of 999 patches (stride = 1), performance actually drops about 1\% from the best mAP. A possible reason is that beyond a certain point, additional granularity introduces redundancy and increases sequence length, making optimization more challenging and reducing performance.
It is worth noting that compared to the conventional square-patch strategy, our method (FFTP) delivers a substantially higher performance across a wide range of patch counts. While the square-patch approach yields only 11.25\% mAP even with over 1,200 patches, FFTP consistently surpasses it, reaching as high as 18.32\% mAP with 991 patches. Even with only 96 patches, FFTP achieves 14.54\% mAP, already outperforming the square-patch result obtained with more than ten times as many patches. It is worth noting that unlike the square patch baseline that exceeds over 1200 patches, the maximum number of patches obtainable under FFTP is 991. This limit arises because reducing the temporal stride to one time-frame produces the densest possible patching configuration, beyond which no further increase in patch count is possible. In Table \ref{tab:efficiency_comparison}, we provide a more detailed analysis of efficiency in terms of FLOPs, training time, and inference latency.

\begin{table}[h]
    \centering
    \scriptsize
    \begin{tabular}{lcccccc} 
        \hline
        Patch Shape & Patches & GFLOPs & Train (hrs) & Inference (ms) & mAP\\
        \hline
        Square & 1212 & 103.35 & 5.3 & 14.50 & 11.25\\
        \hline
        \multirow{4}{*}{FFTP}
               & 96 & {4.15} & 2.3 & {0.96} & 14.54\\
               & 196 & 17.30 & 2.5 & 2.18 & 17.15\\
               & 248 & 21.48 & 2.6 & 2.62 & 17.67\\
               & 496 & 42.79 & 2.8 & 5.38 & 18.01\\
               & \textbf{991} & \textbf{85.32} & \textbf{5.2} & \textbf{11.62} & \textbf{18.32}\\
        \hline
    \end{tabular}
    \vspace{0.5em}
    \caption{Model efficiency across patch counts on AudioSet-18k using AST. All measurements are performed on a single NVIDIA A100 GPU with a fixed audio input length of 10s.}
    \label{tab:efficiency_comparison}
\end{table}

As shown in Table \ref{tab:efficiency_comparison}, the AST model with square patches (16, 16) requires approximately 103.35 GFLOPs per forward pass. In contrast, our most efficient patch configuration, generates only 96 patches, reducing the computational load to just 4.15 GFLOPs while achieving a higher mAP of 14.54. However, the best overall performance is achieved with the FFTP configuration that produces 991 patches. Despite this higher patch count, it remains more efficient than square patching, with a computational load of 85.32 GFLOPs, and achieves the highest mAP of 18.32.

Inference latency is similarly improved: the average latency per sample drops from 14.50 ms with square patches to 0.52 ms with FFTP, enabling faster real-time processing. Even at higher patch counts, latency remains reasonable at 5.38 ms, still well below the square patching baseline. All of our configurations achieve clearly better performance than the square-patch-based AST.

These results confirm that FFTP is not only more accurate but also significantly more efficient in terms of computation, training time, and inference latency, making it well-suited for resource-constrained and real-time audio applications.

\section{Conclusion}
This paper proposes Full-Frequency Temporal Patching (FFTP) for spectrogram-based audio classification models. Through experiments, we demonstrate that FFTP aligns better with the nature of spectrogram data, enhancing the ability of sequence-based models to capture meaningful temporal and spectral information while improving efficiency and accuracy. In addition, we propose SpecMask, a spectrogram-level augmentation technique that structurally masks full frequency bands with localized time-frequency masking, which improves model robustness and complements the representational benefits of FFTP.

\bibliographystyle{IEEEtran}
\bibliography{references}

\end{document}